\documentclass[a4paper,11pt]{article}
\usepackage{jcappub} 
\usepackage{amsfonts}
\usepackage{amsmath}
\usepackage{amssymb}
\usepackage{bm}
\usepackage{float}
\usepackage{array}
\usepackage{makecell}

\usepackage{tikz}
\usetikzlibrary{positioning,arrows.meta,shapes.geometric,calc,backgrounds}

\usepackage{soul}
\usepackage{xcolor}

\newcommand{\HI}{\rm H{\sc i}}

\newcommand{\AAstar}{AA$^{\star}$}

\title{Impact of Residual Gain Errors on the Detectability of Ionized Bubbles during Reionization with SKA 21-cm observations}

\author[a]{Arnab Mishra,}
\author[a,1]{Kanan K. Datta%
\note{Corresponding author},}
\author[a]{Iffat Nasreen,}
\author[a]{Snehasish Saha,}
\author[a]{Sk. Saiyad Ali}
\affiliation[a]{Relativity \& Cosmology Research Centre, Department of Physics, Jadavpur University, Kolkata 700032, India.}

\emailAdd{kanankdatta.physics@jadavpuruniversity.in}

\abstract{Detecting ionized bubbles and characterizing their properties and the surrounding intergalactic medium (IGM) using the \HI\ 21-cm signal provides direct probes of cosmic reionization. Here, we focus on the detection of individual ionized bubbles around luminous sources using a matched-filter technique. We investigate the impact of time- and frequency-correlated residual antenna gain errors on the detectability and characterization of individual ionized regions with SKA-Low. We simulate mock SKA-Low \AAstar\ observations that include the \HI\ 21-cm signal, astrophysical foregrounds, residual gain errors, and system noise corresponding to $100$ h of observing time. We subtract the foreground components from the gain-corrupted data and subsequently apply the matched-filter technique to the residual visibilities. We find that the matched-filter technique is able to detect the ionized bubble and constrain its properties in the presence of residual gain errors with the rms amplitudes up to \(\sigma_A\sim1\%\) and rms phases up to \(\sigma_\phi\sim1^\circ\). Residual gain errors at similar levels prevent the reliable identification of ionized regions in reconstructed \HI\ 21-cm images and bias the estimated \HI\ 21-cm power spectrum. Furthermore, we find that increasing the temporal correlation length increases the level of gain-induced artifacts in the reconstructed image. In contrast, increasing the frequency correlation length reduces these artifacts, as some fraction of the spectrally smooth gain-induced residuals is removed during foreground subtraction. We further find that the inferred bubble and IGM parameters remain largely unaffected by residual gain errors, although their uncertainties can increase. These results demonstrate the potential and robustness of the matched-filter technique for detecting and characterizing individual ionized bubbles in the presence of substantial residual calibration errors.}

\begin{document}
\maketitle
\flushbottom

\section{Introduction}
\label{sec:intro}

The Epoch of Reionization (EoR) marks a landmark episode in cosmic history. During this time, the first luminous sources began to ionize the predominantly neutral intergalactic medium (IGM), over the redshift range $z\sim15$--$6$ \citep{Pritchard_2012,bera2023}. Reionization is expected to have proceeded through the growth of ionized regions around the first galaxies and quasars, which eventually merged as the Universe became fully ionized \citep{mellema2006, Furlanetto_2006, Choudhury_2009}. Recently, several highly luminous quasars and galaxies have been discovered at $z \gtrsim 6$  \cite{Mortlock_2011, Ba_ados_2017, Wang_2018, Wang_2019, Wang_2021, Matsuoka_2019, Matsuoka_2019a, Yang_2020, Witstok2024}. Such sources are expected to produce large ionized regions around them, potentially extending over several tens of comoving Mpc. Detecting these individual ionized bubbles can provide direct information about the size and morphology of ionized regions and the ionization state of the surrounding IGM. In particular, associating these bubbles with known galaxies or quasars can help to establish a direct connection between the properties of the early ionizing sources and their impact on the surrounding IGM. 

The redshifted \HI\ 21-cm signal provides a promising way to probe these large ionized regions. The \HI\ 21-cm emission is nearly zero inside highly ionized regions, while it remains non-zero in the surrounding neutral IGM. This contrast makes it possible, in principle, to identify ionized regions and constrain their physical properties. Tomographic imaging can be used to study their morphology and size distribution \citep{Kakiichi_2017,giri2018}. On the other hand, visibility-based matched filter searches directly in the interferometric data for the expected spatial and spectral signatures of an ionized bubble \citep{Datta_2007, Datta_2008, Datta_2009, Datta_2012, majumdar2012, Ghara_2020, Mishra:2024jjg, Mishra2025, Mahajan:2025pjb}.

However, detecting individual ionized bubbles in the \HI\ 21-cm signal is challenging because of the weak cosmological HI 21-cm signal, bright astrophysical foregrounds, instrumental noise, and other systematics \citep{Ali_2008, 10.1111/j.1365-2966.2012.21889.x}. A matched-filter technique optimally combines the visibility signals from different baselines and frequency channels using a filter matched to the expected spatial and spectral signature of an ionized bubble. This maximizes its detectability. The visibility-based matched-filter approach for detecting individual ionized bubbles was first introduced by \cite{Datta_2007} and subsequently explored for different observational scenarios \citep{Datta_2008,Datta_2009,majumdar2012}. In our previous work, we applied this technique to mock uGMRT and SKA-Low observations, and demonstrated that it can reliably detect individual ionized bubbles even in the presence of strong system noise and foreground contamination \citep{Mishra:2024jjg}. 
We subsequently developed and validated a computationally efficient matched-filter estimator based on gridded visibilities, enabling the analysis of large datasets expected from SKA-Low AA$^\star$ and AA4 observations \citep{Mishra2025}. For the representative ionized regions considered there, ${\rm SNR}\gtrsim10$ could be achieved in approximately $100$~h at $z\sim7$--$8$. We also used a Bayesian MCMC framework to study the prospects of constraining the bubble radius, its three-dimensional location, and the neutral hydrogen fraction of the surrounding IGM. 

However, previous studies assumed perfectly calibrated visibilities. Instrumental calibration introduces an additional challenge for redshifted 21-cm observations. The response of each interferometric element is described by a complex gain, and imperfect estimates of these gains leave residual amplitude and phase errors in the calibrated visibilities. Because the gain errors multiply the total sky signal, even small residual fluctuations can introduce substantial contamination into the data. Spectral structure introduced during calibration is especially problematic for EoR measurements. \cite{Barry2016} showed that an incomplete calibration sky model can move foreground power into modes used to search for the 21-cm signal. \cite{EwallWice2017} further demonstrated that unmodeled foreground sources can introduce frequency-dependent structure in the gain solutions. Similarly, \cite{Trott_Wayth_2016} studied spectral-calibration requirements for SKA-Low and showed that such residual structure can affect both power-spectrum and tomographic measurements. \cite{Aishrila_2022} investigated the impact of residual calibration and source-position errors on both image-plane recovery and the 21-cm power spectrum for SKA1-Low, MWA, and HERA, finding that calibration errors at the $\sim 0.01 \%$ level and that source-position errors of $\gtrsim 5''$ or larger can significantly contaminate the cosmological signal. \cite{Pal_2025} investigated the impact on image-based largest-cluster statistics for simulated SKA-Low observations and found similar results.

Residual calibration errors may remain correlated over neighbouring time samples or frequency channels and therefore need not average down in the same way as independent random fluctuations. \cite{Kumar:2020zja, Kumar:2022lmz} investigated the effect of time-correlated residual gains on 21-cm power-spectrum measurements and showed that they can introduce a systematic bias in the recovered power.
More recently, the effects of residual gains with correlations in both time and frequency have also been investigated, further demonstrating that the correlation properties of the gain errors are important in determining their impact on the recovered 21-cm signal \citep{Gayen:2025kdl}.

In this work, we investigate the impact of residual gain errors on the matched-filter detection of individual ionized bubbles around luminous sources with SKA-Low AA$^\star$. We consider antenna-based complex gains with residual fluctuations in both amplitude and phase. We further consider gain errors that are correlated in both time and frequency. We generate gain-corrupted mock interferometric observations containing the \HI\ 21-cm signal, diffuse Galactic synchrotron emission, and extragalactic point sources, representative of \(100\)~h of SKA-Low AA$^\star$ observations. Our analysis characterizes the robustness of ionized-bubble detection and parameter recovery in the presence of residual-gain errors.

Throughout our analysis, we use the cosmological parameters $ h= 0.7$, $\Omega_{\rm m} = 0.27$, $\Omega_{\Lambda} = 0.73$, $\Omega_{\rm b} = 0.044$, ${\sigma_8=0.83}$~ and ${n_{\rm s}=0.96}$ consistent with the WMAP measurements~\citep{Bennett_2013}.

\section{Mock Observation and Signal Modelling}
\label{sec:simulation}

In this section, we describe the simulations used to generate mock datasets that resemble SKA-Low \AAstar\ observations at frequencies relevant to the epoch of reionization (EoR). These simulations include the cosmological \HI\ 21-cm signal around bright sources and dominant astrophysical foregrounds. The resulting sky maps are converted into visibilities, then sampled  according to the SKA-Low baseline distribution. These visibilities serve as the uncorrupted dataset, upon which instrumental gain errors are introduced in Section~\ref{sec:corrupt_vis}.

\subsection{Observation setup}
\label{subsec:obs_setup}

We consider observations at a central frequency of $175$~MHz, corresponding to a redshift of $z \approx 7.1$ for the \HI\ 21-cm line. This setup is motivated by the discovery of a bright quasar at $z=7.1$~\citep{Mortlock_2011}, which is expected to host a large ionized region. Similar other bright sources and large ionized regions around them are expected to exist during the EoR \citep{Mortlock_2011, Ba_ados_2017, Wang_2018, Wang_2019, Wang_2021, Matsuoka_2019, Matsuoka_2019a, Yang_2020, Witstok2024}.

The simulated sky covers an angular extent of $\sim 3.3^\circ \times 3.3^\circ$, comparable to the primary field of view of SKA-Low at this frequency. The simulation cube consists of $512^3$ grids with a spatial resolution of $1.03$ Mpc (comoving), corresponding to a box size of $527$ Mpc. This yields an angular resolution of $\sim 23^{\prime\prime}$ at $z=7.1$. We use $256$ frequency channels spanning a total bandwidth of $16$ MHz, resulting in a channel width of $\Delta \nu = 62.5$ kHz.

\begin{figure}
\centering
\includegraphics[width=.5\linewidth]{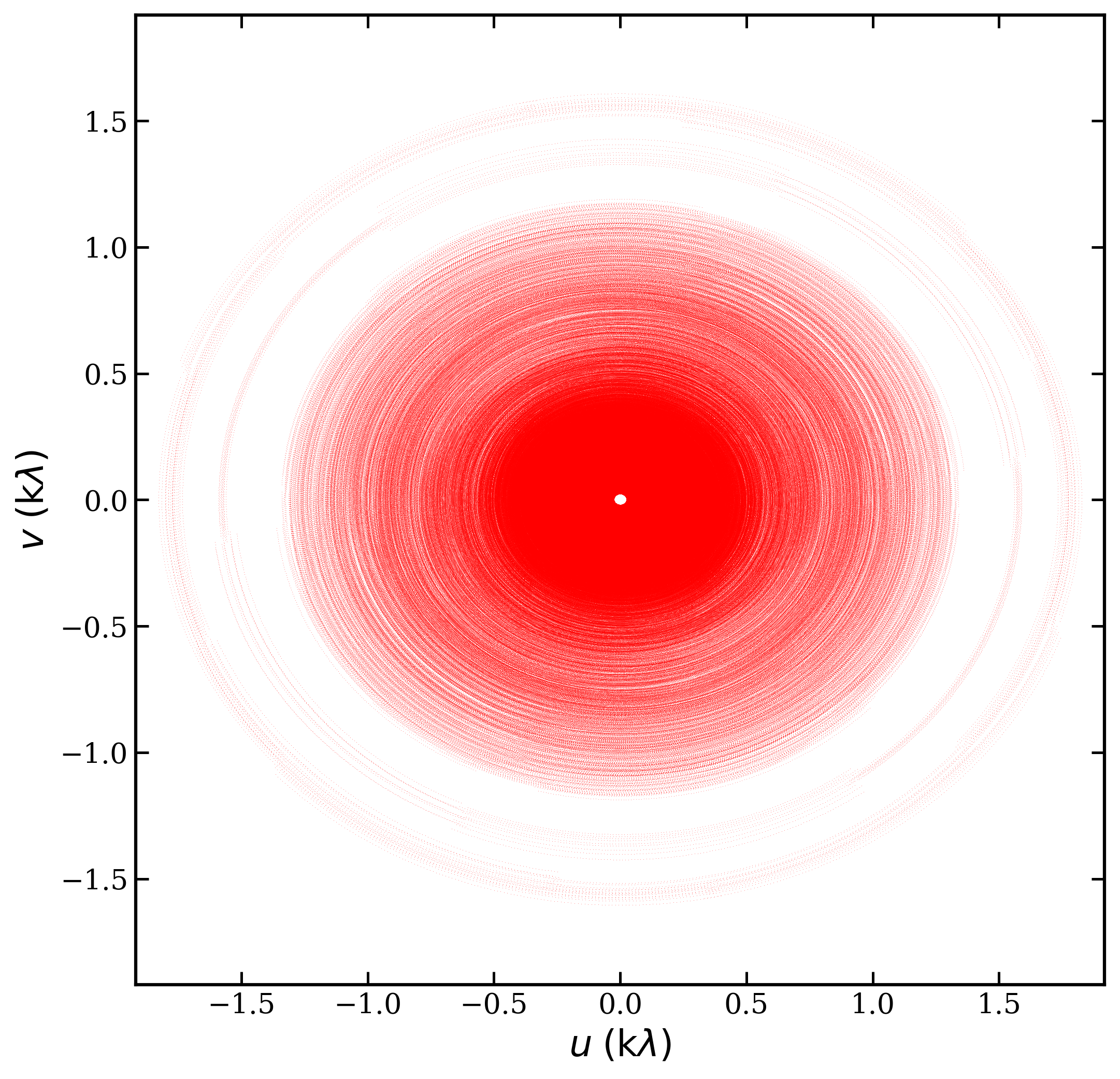}
\caption{$uv$-coverage for $4$ hours of observation at $\nu = 175$ MHz for the SKA-Low \AAstar.}
\label{fig:Baseline_distribution}
\end{figure}

We adopt the SKA-Low \AAstar\ configuration, but restrict our analysis to the central $2$~km radius of the array, which contains $231$ stations. This compact core is densely populated and provides high sensitivity to the large angular scales, suitable for this study. The baseline distribution for $4$ hours of observation, assuming a correlator integration time of $10$ seconds, is shown in Figure~\ref{fig:Baseline_distribution}. The sensitivity parameters ($A_{\rm eff}/T_{\rm sys}$) are taken from the SKA documentation~\citep{skao_performance} as shown in Table~\ref{tab:sensitivity}.

\begin{table}
\centering

\caption{Sensitivity parameters ($A_{\rm eff}/T_{\rm sys}$ in m$^2$ K$^{-1}$) for the SKA-Low station and the \AAstar\ array configurations at the observing frequency of $170\,\rm{MHz}$. Values are taken from the SKA documentation \citep{skao_performance}.}
\label{tab:sensitivity}
\begin{tabular}{c c c c}
\hline
{Frequency (MHz)} & {Single station}  & {\AAstar\ array central 231 stations} \\

\hline
170 & 1.171 & 270.5\\
\hline
\end{tabular}

\end{table}

\subsection{H{\footnotesize I} 21-cm signal}
\label{subsec:HI_sim}

The \HI\ 21-cm signal is simulated following the methodology described in our earlier work~\citep{Mishra:2024jjg}. We generate large-scale dark matter distributions using an $N$-body simulation and identify collapsed halos using a Friends-of-Friends (FoF) algorithm. The simulation volume is $527$ cMpc with $2048^3$ grids, and the minimum halo mass resolved is $\sim 6.26 \times 10^9\, M_\odot$.

Ionization maps and corresponding \HI\ differential brightness temperature cubes are generated using a semi-numerical prescription~\citep{Choudhury_2009, Majumdar_2014, Mondal_2017}. A spherical region of radius $R$ around each grid cell is considered ionized if the condition $\langle n_\gamma \rangle_R \geq \langle n_{\rm H} \rangle_R$ is satisfied. Here, $\left\langle n_{\gamma}\right\rangle_R$ and $\left\langle n_{\rm H}\right\rangle_R$ denote the average number densities of ionizing photons and hydrogen atoms within a sphere of radius $R$. Cells that do not satisfy this criterion are assigned a partial ionized fraction following the same prescription. The resulting ionization and brightness temperature cubes are subsequently downsampled from a resolution of $2048^3$ to $512^3$ cells while preserving the same comoving box size of $527$ cMpc. This downsampled cube is used for all subsequent visibility simulations and analysis.

In this work, we consider a representative scenario corresponding to an ionized bubble embedded in a partially neutral intergalactic medium with a mean neutral fraction $x_{\rm HI} = 0.88$ at $z=7.1$. The ionized bubble is created by a bright QSO-like source, similar to the one found by \cite{Mortlock_2011}. We refer to our previous work \citep{Mishra:2024jjg} for more details.  This controlled setup allows us to investigate the impact of instrumental gain errors on the detectability of ionized bubbles.

\subsection{Foregrounds}
\label{subsec:Foregrounds}

The simulated sky also includes dominant astrophysical foregrounds, which are several orders of magnitude brighter than the cosmological \HI\ 21-cm signal. We consider two major foreground components: diffuse galactic synchrotron emission (DGSE) and extragalactic point sources. 

The DGSE is modeled as a statistically isotropic Gaussian random field with an angular power spectrum given by~\citep{Choudhuri_2014}:
\begin{equation}
C^M_l(\nu) = A_{150} \left(\frac{1000}{l}\right)^{\beta} \left(\frac{\nu_0}{\nu}\right)^{2\alpha},
\end{equation}
where $\beta = 2.34$, $\alpha = 2.8$, and $A_{150} = 513\, {\rm mK}^2$.

The extragalactic point sources are generated using the observed differential source count distribution at $150$ MHz \citep{10.1111/j.1365-2966.2012.21889.x}: 
\begin{equation}
\frac{dN}{dS} = \frac{10^{3.75}}{\text{Jy\,sr}} \left(\frac{S}{1\,\text{Jy}}\right)^{-1.6}.
\end{equation}
We consider sources with flux densities in the range $0.1$--$1000$ mJy and assign spectral indices randomly between $0.7$ and $0.8$. We follow the approach adopted in our previous work~\citep{Mishra:2024jjg} to simulate these two foreground components.

The \HI\ signal and foreground maps are combined to generate gridded multi-frequency sky cubes, which are subsequently Fourier transformed to obtain visibilities at gridded baselines. Subsequently, mock visibilities are generated at all SKA-Low \AAstar\ baselines using the nearest grid point (NGP) method. We emphasize that the visibilities generated in this section correspond to the uncorrupted dataset. Instrumental gain errors are subsequently introduced at the visibility level, as described in subsection~\ref{subsec:gain_errors}.

\subsection{Noise contribution}
\label{subsec:Noise}

The system noise contribution to each measured visibility is modeled as a Gaussian random variable with zero mean. The root mean square (rms) noise for each baseline and frequency channel is given by~\citep{Datta_2007}:
\begin{equation}
\sqrt{\langle N^2 \rangle} = \frac{\sqrt{2} k_B T_{\rm sys}}{A_{\rm eff} \sqrt{\Delta\nu \Delta t}},
\label{eq:sys_noise}
\end{equation}
where $k_B$ is the Boltzmann constant, $T_{\rm sys}/A_{\rm eff}$ is the system-equivalent flux density, $\Delta\nu$ is the channel width, and $\Delta t$ is the correlator integration time. For $\Delta\nu = 62.5$ kHz and $\Delta t = 10$ sec, the rms noise per visibility is $\sim{2.11} $ Jy at $175$ MHz. The noise described here is used in the gridding pipeline (subsection~\ref{subsec:gridding}). There, the noise is appropriately scaled to represent longer observing times.

\section{Simulation of Corrupted Visibilities}
\label{sec:corrupt_vis}

\subsection{Modeling Instrumental Gain Errors}
\label{subsec:gain_errors}
In radio interferometric observations, the incoming radio signal is affected by the Earth's ionosphere and different components of the receiving system, including antennas, receivers, and amplifiers. These effects are described by complex antenna gains, which can vary with time and frequency. Although calibration is used to correct these gains, they cannot be estimated perfectly, leaving residual errors in their amplitude and phase. These residual gain errors affect both the astrophysical foregrounds and the \HI\ 21-cm signal. Since the foregrounds are much brighter than the cosmological signal, even small residual gain errors can introduce significant contamination, making the detection of the weak redshifted \HI\ 21-cm signal more challenging \citep{Thompson2017,Gayen:2025kdl}. In this work, we model the residual gain errors as direction-independent, antenna-based stochastic fluctuations in amplitude and phase, with correlations in time, frequency, or both. In the presence of residual gain errors and system noise, the observed visibility between a pair of antennas $i$ and $j$ can be written as

\begin{equation}
V_{ij}^{\rm obs}(t,\nu) = g_i(t,\nu)\, g_j^*(t,\nu)\, V_{ij}^{\rm true}(t,\nu) + N_{ij}(t,\nu),
\end{equation}
where $V_{ij}^{\rm true}(t,\nu)$ denotes the true sky visibility, $g_i(t,\nu)$ is the complex gain associated with the $i$-th antenna, and $N_{ij}(t,\nu)$ represents the system noise contribution. The noise is assumed to be a Gaussian random variable with zero mean and is uncorrelated across different baselines and frequency channels \citep{Thompson2017}.

The antenna-based complex gain can be expressed in terms of amplitude and phase perturbations as
\begin{equation}
g_i(t,\nu) = \left[1 + \Delta A_i(t,\nu)\right] \exp\left[i\,\Delta \phi_i(t,\nu)\right],
\end{equation}
where $\Delta A_i(t,\nu)$ and $\Delta \phi_i(t,\nu)$ denote the fractional amplitude and phase errors, respectively. In the absence of calibration errors, $\Delta A_i = 0$ and $\Delta \phi_i = 0$, reducing the observed visibility to the true visibility. In the present analysis, the residual gains are assumed to be direction-independent. Thus, $g_i(t,\nu)$ depends on antenna, time, and frequency, but not on sky direction. Direction-dependent effects, such as errors in the station primary beam or ionospheric calibration, are not included in the present gain model.

Following earlier works \citep{Gayen:2025kdl,Gayen:2024vaq}, we assume the residual gain errors as samples drawn from a stochastic process characterized by its variance and correlation function. We also assume that the gain errors are antenna-based, statistically identical for all antennas, and uncorrelated between different antennas. We model the correlation of gain errors using Gaussian two-point correlation functions. The temporal correlation of the gain fluctuations is described by
\begin{equation}
\eta(\tau) = \exp\left(-\frac{\tau^2}{2T_{\rm corr}^2}\right),
\label{eq:corr_time_eq}
\end{equation}
where $T_{\rm corr}$ is the characteristic time-correlation scale. We adopt $T_{\rm corr}=20$~s, which is larger than the correlator integration time of $10$~s and therefore allows the residual gain errors to remain correlated across neighbouring time samples. We also studied larger time correlation lengths to examine the sensitivity of our results, which has been discussed in Section~\ref{sec:results}.

In addition to time dependence, residual calibration errors also exhibit frequency structure due to imperfect bandpass calibration. The frequency correlation of  gain errors (bandpass response) can also be modeled using a two-point correlation function as \citep{Gayen:2025kdl},
\begin{equation}
\xi(\Delta \nu) = \exp\left(-\frac{\Delta \nu^2}{2\nu_{\rm corr}^2}\right),
\label{eq:corr_freq_eq}
\end{equation}
where $\nu_{\rm corr}$ is the characteristic frequency correlation scale. We adopt $\nu_{\rm corr}=1$~MHz, which is much larger than the channel width of $62.5$~kHz and therefore produces gain fluctuations that remain correlated across several neighbouring frequency channels. Similarly, here we also studied larger frequency correlation lengths to examine the sensitivity of our results, which has been discussed in Section~\ref{sec:results}.

\subsection{Generation of Correlated Gain Realizations}
\label{subsec:gain_generation}

To generate the residual gain errors, we employ Gaussian Process (GP) sampling with covariance matrices constructed from the prescribed temporal or spectral correlation functions. We first consider time- and frequency-correlated gain errors separately, by introducing correlations along either the time or the frequency axis while keeping the other axis statistically independent. We then also consider a joint time--frequency-correlated model in which the gain fluctuations are correlated simultaneously along both axes.

For each antenna, we first generate an uncorrelated Gaussian random matrix $\mathbf{Z} \sim \mathcal{N}(0,1)$ of dimensions $N_t \times N_\nu$, where $N_t$ and $N_\nu$ denote the number of time samples and frequency channels, respectively. The covariance matrix $\mathbf{C}_{\rm time}$ of dimensions $N_t \times N_t$ corresponding to the temporal correlation function defined in Eqs.~(\ref{eq:corr_time_eq}) is first constructed. Similarly, covariance matrix  $\mathbf{C}_{\rm freq}$ of dimensions $N_{\nu} \times N_{\nu}$ corresponding to spectral correlation function is constructed using Eqs.~(\ref{eq:corr_freq_eq}). To generate Gaussian random fields with the desired correlation properties, these covariance matrices are decomposed using the Cholesky decomposition as \citep{RasmussenWilliams2005},
\begin{equation}
\mathbf{C}_{\rm time}=\mathbf{L}_{\rm time}\mathbf{L}_{\rm time}^{\rm T},
\end{equation}
for the temporal covariance matrix, and
\begin{equation}
\mathbf{C}_{\rm freq}=\mathbf{L}_{\rm freq}\mathbf{L}_{\rm freq}^{\rm T},
\end{equation}
for the spectral covariance matrix, where $\mathbf{L}_{\rm time}$ and $\mathbf{L}_{\rm freq}$ are lower triangular matrices with dimensions same as the corresponding covariance matrices.

For the time-correlated gain model, the temporal correlations are introduced by multiplying the  uncorrelated Gaussian random matrix $\mathbf{Z}$  with the Cholesky factor of the temporal covariance matrix $\mathbf{L}_{\rm time}$ as,
\begin{equation}
\mathbf{X}=\mathbf{L}_{\rm time}\mathbf{Z},
\end{equation}
where $\mathbf{X}$ denotes the correlated gain field. This operation introduces the required temporal correlation in antenna-based gain errors while preserving the statistical independence between different frequency channels. Similarly, for the frequency-correlated gain model, correlations are introduced along the frequency axis through
\begin{equation}
\mathbf{X}=\mathbf{Z}\mathbf{L}_{\rm freq}^{\rm T},
\end{equation}
which produces gain fluctuations with the desired spectral correlation while keeping different time samples statistically independent. 

In addition to the two limiting cases described above, we also consider gain fluctuations that are correlated simultaneously in time and frequency. Assuming separable temporal and spectral correlations, the correlated Gaussian field is generated as

\begin{equation}
    \mathbf{X} = \mathbf{L}_{\rm time}\, \mathbf{Z}\, \mathbf{L}_{\rm freq}^{T}.
    \label{eq:joint_corr}
\end{equation}

The left and right multiplications introduce the required temporal and frequency correlations, respectively. The resulting field is normalized to unit rms and scaled by the corresponding gain-error amplitude, $\sigma_A$ or $\sigma_\phi$. Independent realizations are generated for the amplitude and phase fluctuations of each antenna.

For simplicity, we assume that the residual amplitude and phase fluctuations have the same temporal and frequency correlation scales, i.e., $T_{\rm corr}^{A}=T_{\rm corr}^{\phi}=T_{\rm corr}$ and $\nu_{\rm corr}^{A}=\nu_{\rm corr}^{\phi}=\nu_{\rm corr}$, while their stochastic realizations are generated independently. In practice, these correlation scales may differ depending on the physical origin of the residual gain errors.

\subsection{Validation of the Gain Simulations}

\begin{figure}
\centering
\includegraphics[width=\linewidth]{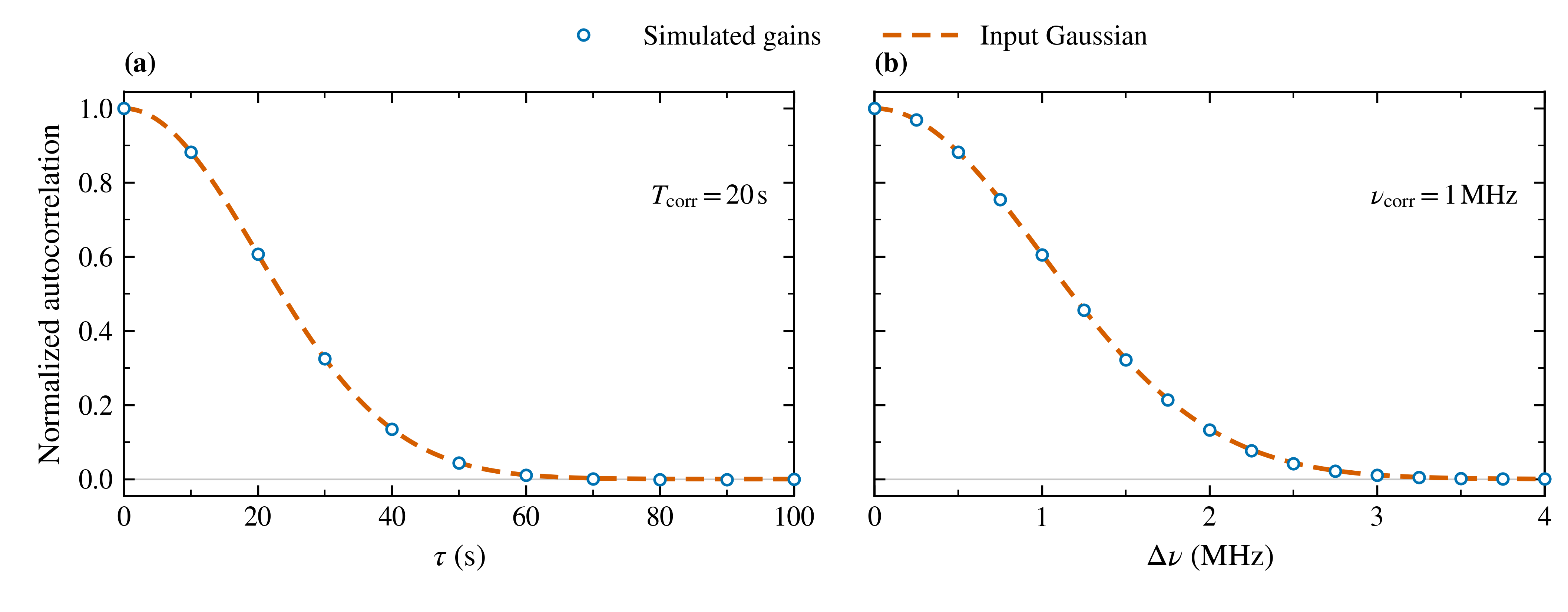}
\caption{
Normalized autocorrelation functions of the simulated residual gain-amplitude fluctuations (open circles) compared with the input Gaussian correlation models (dashed curves). Panel (a) shows the temporal correlation for $T_{\rm corr}=20\,{\rm s}$, while panel (b) shows the frequency
correlation for $\nu_{\rm corr}=1\,{\rm MHz}$. The estimates are averaged over antennas  
with a global mean and variance used for normalization. 
The close agreement demonstrates that the generated gain realizations reproduce the prescribed correlation structure. The phase fluctuations show similar agreement and are not shown here.
}
\label{fig:gain_correlation_validation}
\end{figure}

To validate the implementation, we estimate the normalized autocorrelation functions of the simulated gain amplitudes along both the time and frequency axes and compare them with the corresponding input Gaussian correlation models. Figure~\ref{fig:gain_correlation_validation} shows the comparison for a representative realization with $T_{\rm corr}=20\,{\rm s}$ and $\nu_{\rm corr}=1\,{\rm MHz}$. The measured correlations closely follow the input models, confirming that the generated gain realizations reproduce the prescribed correlation properties. We performed the same validation for the phase fluctuations and found similarly good agreement with the corresponding input correlation models.

To separately study the impact of different correlation properties of gain errors on the detectability of the ionized bubbles in the \HI\ 21-cm signal, we consider two limiting cases:
(i) time-correlated gain errors, where gain errors are correlated over nearby time steps  with no correlation between different frequency channels, and
(ii) frequency-correlated gain errors, where errors are correlated over nearby frequency channels with no correlation between different time steps.
 In addition to these two limiting cases, we also consider the joint time--frequency-correlated model defined in Eq.~\ref{eq:joint_corr}, in which the gain fluctuations are correlated simultaneously along both axes. In all three cases, the gain errors are assumed to be uncorrelated between different antennas.

\subsection{Gridding of Gain-Corrupted Visibilities and system Noise Incorporation}
\label{subsec:gridding}

The simulated visibilities at all SKA-Low \AAstar\ baselines, described in Section~\ref{sec:simulation}, are corrupted using the complex antenna-based gain model presented in Section~\ref{subsec:gain_errors}. Here, the true sky visibility includes the combined contributions from the \HI\ 21-cm signal, diffuse Galactic synchrotron emission, and extragalactic point sources before the instrumental gain corruption is applied. The gain corruption is applied through the corresponding complex gain products of the two antennas forming each baseline. The treatment adopted for the effective $100$~h observation is described below.

The simulated dataset corresponds to $4$ hours of observation with a correlator integration time of $10$~s, resulting in $1440$ time samples. We consider $231$ SKA-Low AA$^\star$ stations within the central $2$ km radius, which provide $26565$ unique baselines at each time step. With $256$ frequency channels, the full dataset therefore contains $\sim 10^{10}$ complex visibility samples across baseline, time, and frequency. The total number of complex visibilities could be $2$ to $3$ orders of magnitude larger in realistic observational scenarios. The direct application of the foreground subtraction and matched-filter techniques to such a large, irregularly sampled visibility data set is computationally expensive. We therefore employ the fast visibility gridding algorithm developed in our previous work \citep{Mishra2025}, which maps the irregularly sampled visibilities onto a regularly spaced three-dimensional $(u,v,\nu)$ grid.

The gridding is performed independently for each frequency channel on a $512\times512$ uniform $uv$ grid, producing a final visibility cube of dimensions $256\times512\times512$. For each grid cell, all visibilities falling within the corresponding $(u,v)$ bin are summed and divided by the number of contributing visibility samples to that grid cell. This averaging reduces the data volume substantially while retaining the visibility information for the subsequent foreground subtraction and matched filtering analysis. This gridded representation substantially reduces the computational cost of foreground subtraction and matched-filter analysis. At the same time, it preserves the spatial and spectral signatures of the \HI\ 21-cm signal around ionized bubbles, which are essential for their detection \citep{Mishra2025}.

To represent the residual gain errors for a total observing time of $100$~h, we generate $25$ independent gain-error realizations, each corresponding to a $4$~h observing session. The rms amplitude and phase errors and the adopted correlation properties are kept fixed for all realizations. We assume that the same $4$~h LST range is observed in each session. Therefore, the underlying sky visibilities and the corresponding $uv$ sampling are identical between sessions, while the residual gain realization is generated independently for each session. For each baseline, the corresponding gain products are averaged over the $25$ independent realizations, and the averaged baseline gain product is then applied to the simulated visibility. Under the repeated-LST-track assumption, this is equivalent to averaging the $25$ independently gain-corrupted visibility datasets. Thus, the rms gain errors of each individual $4$~h session remain unchanged, while the independent residual gain fluctuations partially average down when the sessions are combined. The resulting gain-corrupted visibilities are then gridded following the procedure described above.

After gridding the gain-corrupted visibilities, system noise is incorporated into the visibility cube. The statistical properties of the system noise are described in Section~\ref{subsec:Noise}, where it is modeled as an independent Gaussian random variable with zero mean and variance determined by the system temperature, effective collecting area, observing bandwidth, and integration time. Since multiple visibility samples contribute to a given $(u,v)$ grid cell, the noise rms is reduced through averaging. The noise rms in each grid cell is therefore given by $\sigma_{\rm grid}=\sigma_{\rm vis}/\sqrt{n_b}$, where $\sigma_{\rm vis}$ is the rms noise of an individual visibility and $n_b$ denotes the number of independent visibility samples contributing to that grid cell.

The simulated visibility data correspond to 4 hours of observation with a 10~s correlator integration time. To represent a total observing time of $100$~h, corresponding to 25 such $4$~h observing sessions, the system-noise rms is further reduced by a factor of $\sqrt{25}$. 
The system noise realization is generated independently for each grid cell and frequency channel and is subsequently added to the gain-corrupted gridded visibilities. The resulting visibility cube therefore contains the combined effects of the cosmological 21-cm signal, astrophysical foregrounds,  residual calibration errors, and realistic system noise. This noisy and corrupted visibility cube serves as the input for the foreground subtraction stage described in the following subsection.

\begin{figure*}
\centering
\resizebox{\textwidth}{!}{%
\begin{tikzpicture}[
  >={Stealth[length=2mm,width=1.35mm]},
  font=\fontsize{8.5}{10.2}\selectfont,
  box/.style={draw=black!65,line width=0.55pt,rounded corners=2pt,
              fill=white,align=center,text width=3.50cm,
              inner xsep=5.5pt,inner ysep=4pt,minimum height=1.16cm},
  smallbox/.style={box,minimum height=0.78cm},
  compactbox/.style={box,text width=3.20cm},
  header/.style={anchor=west,font=\fontsize{12}{14}\selectfont\bfseries,
                 inner sep=0pt},
  flow/.style={->,draw=black,line width=0.65pt,rounded corners=2pt},
  mergefeed/.style={draw=black,line width=0.6pt,rounded corners=2pt},
  feed/.style={->,draw=black,line width=0.55pt}
]
\begin{scope}[on background layer]
  \filldraw[fill=white,draw=black!35,rounded corners=3pt,line width=0.45pt]
    (-0.54,0.50) rectangle (17.38,-2.55);
  \filldraw[fill=white,draw=black!35,rounded corners=3pt,line width=0.45pt]
    (-0.54,-2.85) rectangle (17.38,-6.80);
  \filldraw[fill=white,draw=black!35,rounded corners=3pt,line width=0.45pt]
    (-0.54,-7.11) rectangle (17.38,-10.52);
\end{scope}

\node[header] at (-0.28,0.00) {\quad Sky model and visibility generation};
\node[smallbox] (hi) at (2.00,-0.85)
  {\textbf{H\textsc{i} 21-cm 3D sky signal}};
\node[smallbox] (fg) at (2.00,-1.85)
  {\textbf{Foregrounds}\\ DGSE and point sources};
\node[box] (fft) at (6.45,-1.35)
  {\textbf{Apply FFT}\\ Gridded sky visibilities};
\node[box] (sample) at (10.90,-1.35)
  {\textbf{True visibilities}};
\node[box] (array) at (15.35,-1.35)
  {\textbf{SKA-Low}\\ \textbf{configuration}\\ Antenna and baseline\\ distributions};

\coordinate (skyjoin) at (4.20,-1.35);
\draw[mergefeed] (hi.east) -- (skyjoin) -- (fg.east);
\draw[flow] (skyjoin) -- (fft.west);
\draw[flow] (fft.east) -- (sample.west);

\node[header] at (-0.28,-3.31)
  {\quad Corrupted visibility generation and gridding};
\node[box] (gains) at (15.35,-4.40)
  {\textbf{Residual station gains}\\ Time / frequency / joint\\
   $g_i=(1+\Delta A_i)e^{\mathrm{i}\Delta\phi_i}$};
\node[compactbox] (corrupt) at (7.15,-5.05)
  {\textbf{Corrupted visibility}\\ \textbf{generation}};
\node[box] (grid) at (2.05,-4.40)
  {\textbf{Apply fast estimator}\\ Gridded visibilities\\ generated};
\node[smallbox] (noise) at (2.05,-6.00)
  {\textbf{System noise}};

\coordinate (gainjoin) at (10.90,-5.05);

\draw[mergefeed] (sample.south) -- (gainjoin);

\draw[mergefeed] (array.south) -- (15.35,-2.69)
  -- (12.10,-2.69) -- (12.10,-4.05) -- (gainjoin);

\draw[mergefeed] (gains.west) -- (12.48,-4.40)
  -- (12.48,-5.05) -- (gainjoin);

\fill[black] (gainjoin) circle[radius=1.4pt];
\draw[flow] (gainjoin) -- (corrupt.east);

\draw[flow] (corrupt.west) -- (4.72,-5.05)
  -- (4.72,-4.40) -- (grid.east);
\draw[feed] (noise.north) -- (grid.south);

\node[header] at (-0.28,-7.54)
  {\quad Foreground subtraction and analysis};
\node[box] (gpr) at (2.00,-8.95)
  {\textbf{GPR foreground}\\ \textbf{subtraction}\\ Residual visibilities};
\node[box] (mf) at (8.675,-8.95)
  {\textbf{Apply matched filter}\\ Estimate SNR and\\ bubble radius};
\node[box] (mcmc) at (15.35,-8.95)
  {\textbf{Bayesian inference}\\ MCMC posteriors\\
   $R_f,\,\theta_x,\,\theta_y,\,\Delta\nu,\,x_{\mathrm{HI}}$};

\draw[flow] (grid.west) -- (-0.32,-4.40)
  -- (-0.32,-8.95) -- (gpr.west);
\draw[flow] (gpr.east) -- (mf.west);
\draw[flow] (gpr.north) -- (2.00,-8.05)
  -- (15.35,-8.05) -- (mcmc.north);
\end{tikzpicture}%
}
\caption{Flowchart illustrating the mock-observation and analysis pipeline.}
\label{fig:pipeline}
\end{figure*}
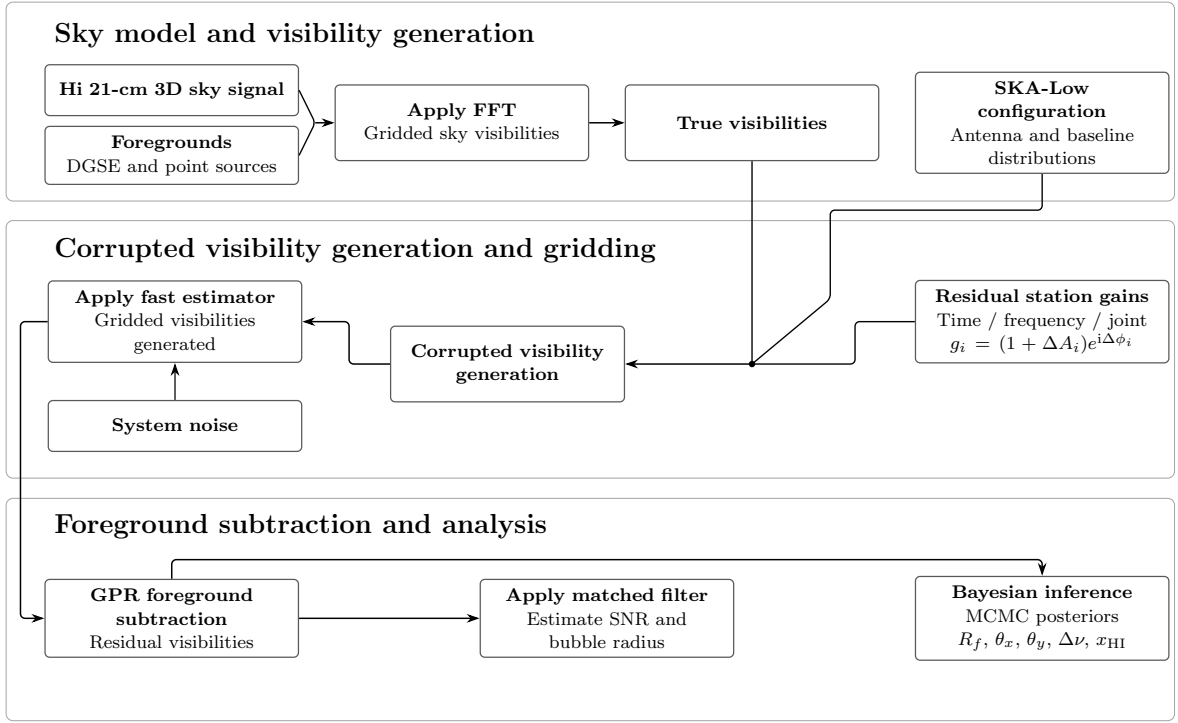

\section{Foreground Subtraction using Gaussian Process Regression}
\label{sec:gpr}
Following the addition of system noise, we apply Gaussian Process Regression (GPR) \citep{Mertens2018} to the gain-corrupted gridded visibility cube to model and subtract the astrophysical foreground contribution. The visibility cube contains the \HI\ 21-cm signal together with diffuse Galactic synchrotron emission (DGSE), extragalactic point sources, residual gain errors, and system noise. 

The foreground subtraction is performed independently along the frequency axis for each grid cell of the gridded visibility cube. We do not model the DGSE and point-source contributions separately. Instead, GPR is applied to their combined spectral contribution in each $(u,v)$ grid cell. Since both foreground components are expected to vary smoothly with frequency compared with the \HI\ 21-cm signal, we model the foreground covariance using a radial basis function (RBF) kernel \citep{Mertens2018,Ghosh2020} of the form

\begin{equation}
K(\nu,\nu')=\sigma_f^2\exp\left[-\frac{(\nu-\nu')^2}{2l^2} \right],
\end{equation}
where $\sigma_f^2$ denotes the foreground variance and $l$ is the characteristic spectral correlation length that determines the smoothness of the foreground emission. The reconstructed foreground model is then subtracted from the gridded visibility cube. The resulting residual visibilities are used for the matched-filter analysis described in the following subsection.

The implementation of the GPR algorithm follows our previous work \citep{Mishra2025}. In this work, we adopt GPR kernel correlation lengths of $l = 12\,\mathrm{MHz}$.

\section{Matched Filter}
\label{sec:MF}

Following the foreground subtraction, the residual visibility cube contains the HI 21-cm signal together with the system noise and any residual instrumental calibration errors arising from imperfect gain correction. To enhance the detectability of the weak HI signal, we employ the visibility-based matched filtering technique originally proposed by \cite{Datta_2007} and subsequently explored in several studies of ionized bubble detection \citep{Datta_2008, Datta_2012,majumdar2012, Mishra:2024jjg, Mishra2025}. The matched filter is constructed using the expected visibility signature of a spherical ionized bubble which is buried in HI medium and provides the maximum signal-to-noise ratio (SNR) when the filter matches the target signal. The matched-filter estimator is given by,
\begin{equation}
    {\hat{E}} = \frac{\sum_{a,b}  S_f^*(\mathbf{U}_a,\nu_b)  \hat{V}(\mathbf{U}_a,\nu_b)}  {\sum_{a,b}1},
    \label{eq:estim}
\end{equation}
where $S_f(\mathbf{U},\nu)$ is the matched filter corresponding to a bubble of a given radius, and $\hat{V}(\mathbf{U},\nu)$ denotes the residual visibility after foreground subtraction. The summation extends over all baselines and frequency channels.

Assuming that the system noise in different baselines and frequency channels is uncorrelated and Gaussian distributed, the variance of the estimator is

\begin{equation}
   \left<(\Delta {\hat{E}})^2\right>_{\rm NS} = {\langle N^2\rangle} \, \frac{\sum_{a,b} \left|S_f(\mathbf{U}_a,\nu_b)\right|^2} {\left(\sum_{a,b}1\right)^2},
\end{equation}
where $\langle N^2\rangle$ is the variance of the system noise for an individual visibility, computed using Eq.~(\ref{eq:sys_noise}). The corresponding signal-to-noise ratio is

\begin{equation}
{\rm SNR}=\frac{\langle \hat{E}\rangle}{\sqrt{\left<(\Delta {\hat{E}})^2\right>_{\rm NS}}},
\label{eq:SNR}
\end{equation}
which is evaluated as a function of the filter radius. The peak of the SNR curve provides both the detection significance and the
recovered bubble radius. Throughout this work, we adopt ${\rm SNR}\ge5$ as the criterion for a robust detection.

\begin{figure}
    \centering
    \includegraphics[width=0.9\linewidth]{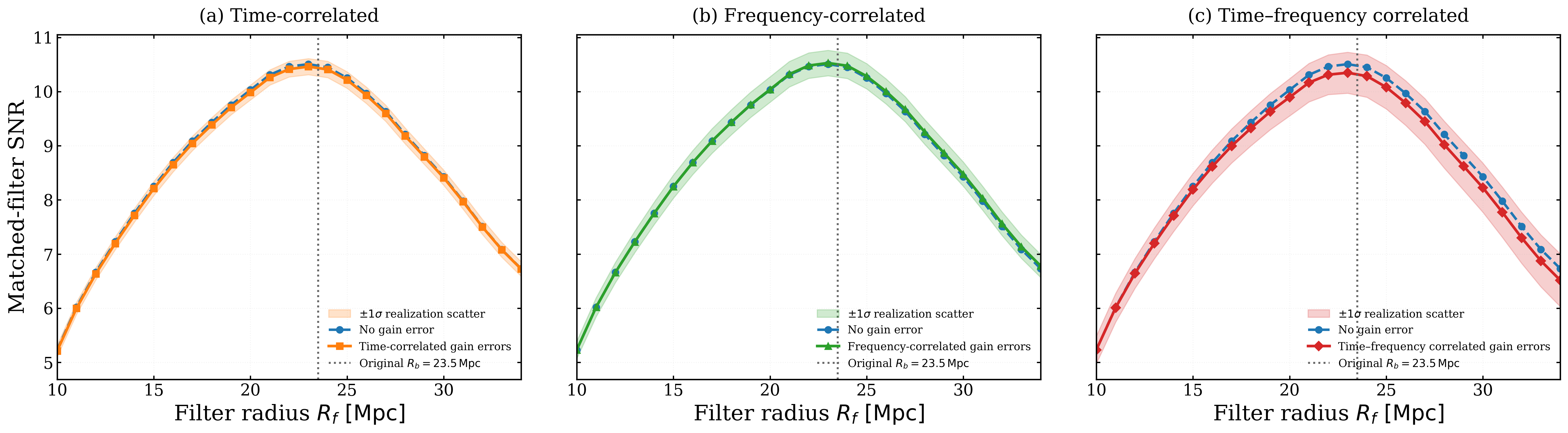}
    \caption{Mean matched-filter SNR as a function of filter radius for the $100$~h of SKA-Low \AAstar observations. The three panels show the results for (a) time-correlated gain errors with $T_{\rm corr}=20$~s, (b) frequency-correlated gain errors with $\nu_{\rm corr}=1$~MHz, and (c) joint time- and frequency-correlated gain errors with $T_{\rm corr}=20$~s and $\nu_{\rm corr}=1$~MHz. The no-gain result is shown for comparison. The solid curves show the mean over ten independent gain realizations, while the shaded regions show the corresponding $1\sigma$ scatter.}
    \label{fig:mf_gain}
\end{figure}

\begin{figure}
    \centering
    \includegraphics[width=\linewidth]{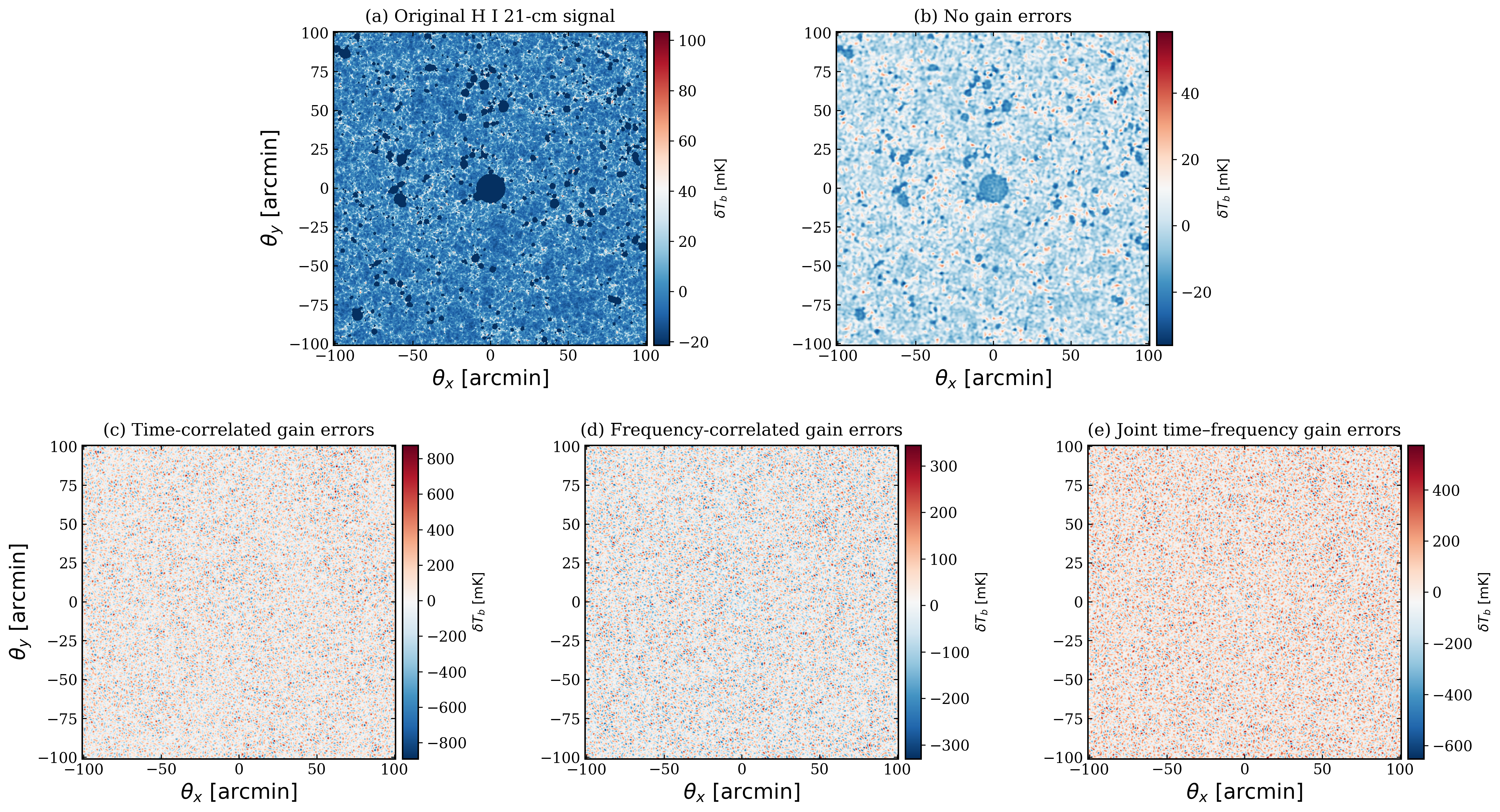}
   \caption{Central-frequency-channel brightness-temperature maps for the $100$~h observation without system noise. From top left to bottom right, the panels show (a) the original mean-subtracted \HI\ 21-cm brightness-temperature field, (b) the reconstructed image after foreground subtraction without residual gain errors, (c) the reconstructed image for the time-correlated gain-error model with $T_{\rm corr}=20$~s, (d) the reconstructed image for the frequency-correlated gain-error model with $\nu_{\rm corr}=1$~MHz, and (e) the reconstructed image for the joint time--frequency-correlated gain-error model with $T_{\rm corr}=20$~s and $\nu_{\rm corr}=1$~MHz.}
    \label{fig:imaging_100h_gainonly}
\end{figure}

\begin{figure}
    \centering
    \includegraphics[width=0.6\linewidth]{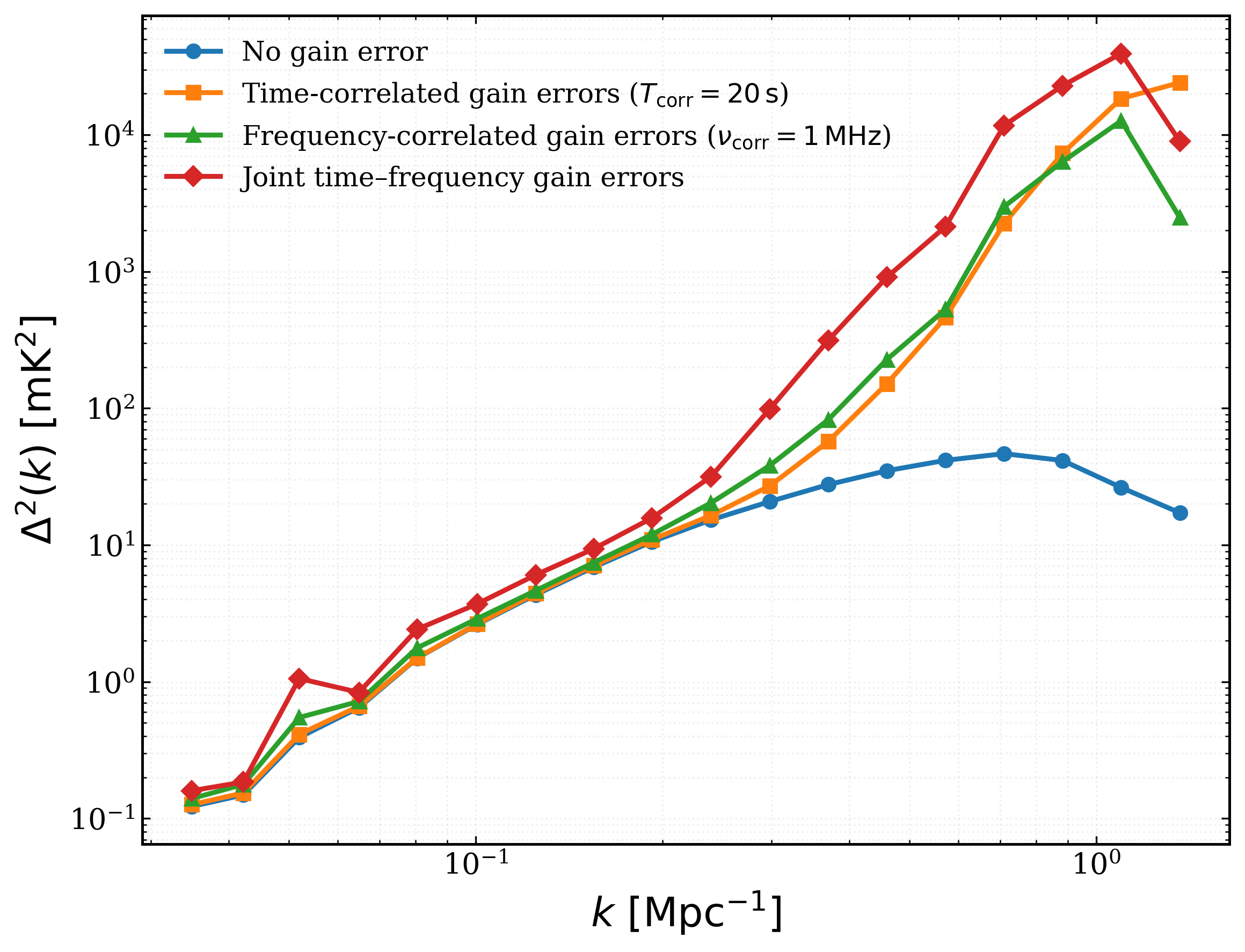}
    \caption{Spherically averaged 21-cm power spectrum, $\Delta^2(k)$, recovered from the foreground-subtracted brightness-temperature cube for the $100$~h observation. The no-gain reference is compared with the time-correlated ($T_{\rm corr}=20$~s), frequency-correlated ($\nu_{\rm corr}=1$~MHz), and joint time-frequency-correlated ($T_{\rm corr}=20$~s$,\ \nu_{\rm corr}=1$~MHz) residual-gain models. For all gain-error cases, the rms gain fluctuations are $\sigma_A=1\%$ and $\sigma_\phi=1^\circ$. System noise is not included here.}
    \label{fig:delta2_100h_gain}
\end{figure}

\section{Results}
\label{sec:results}

The results presented in this section correspond to a simulated ionized bubble of radius $R_b \approx 23.5$~Mpc at $z=7.1$. The bubble is embedded in an IGM with a neutral hydrogen fraction of $x_{\rm HI}=0.88$.
In the gain-corrupted simulations, each $4$-h observing session is assumed to represent an independent realization of the antenna-based residual gains, with the rms amplitude and phase fluctuations fixed at $\sigma_A=1\% $ and $\sigma_\phi=1^\circ$, respectively.
We consider three gain-correlation models: a time-correlated model with $T_{\rm corr}=20$~s, a frequency-correlated model with $\nu_{\rm corr}=1$~MHz, and a joint time-frequency-correlated model with $T_{\rm corr}=20$~s and $\nu_{\rm corr}=1$~MHz.

\subsection{Robustness of matched-filter bubble detection in presence of residual gain error}
\label{subsec:mf_results}
We first investigate how residual gain errors affect the matched-filter detection of the ionized bubble. Figure~\ref{fig:mf_gain} shows the matched-filter SNR as a function of the filter radius, $R_f$, for the three gain-error models considered here. The left, middle, and right panels show the time-correlated, frequency-correlated, and joint time- and frequency-correlated cases, respectively. In each panel, the no-gain result is shown for comparison. For the gain-error cases, the solid curve represents the mean over ten independent residual gain realizations, while the shaded region shows the corresponding $1\sigma$ realization-to-realization scatter. The same system-noise realization is used for all ten realizations, while the residual gain errors are generated independently.  

For the reference case without residual gain errors, the matched-filter SNR reaches a maximum of approximately $10.5$ at $R_f\simeq23$~Mpc, close to the characteristic size of the actual ionized region. The time-correlated gain-error case gives a very similar mean SNR profile. Individual gain realizations produce some variation, but the position of the SNR peak changes only slightly and remains close to that of the reference case. Thus, for the fiducial values $\sigma_A=1\%$, $\sigma_\phi=1^\circ$, and $T_{\rm corr}=20$~s, the recovered bubble scale is only weakly affected by time-correlated residual gain errors.

A similar trend is seen for the frequency-correlated model in the middle panel of Figure~\ref{fig:mf_gain}. The mean SNR again peaks close to the actual bubble radius and follows the no-gain-error result reasonably well. We observe a small shift in the mean SNR towards higher values. The peak position remains very close to the actual bubble radius across different realizations. However, we find that the primary effect of the frequency-correlated residual gain errors is to increase the realization-to-realization variation, rather than to introduce any systematic shift in the recovered bubble radius. The small increase in the mean SNR is not statistically significant, as it remains within the uncertainty range.

The right panel of Figure~\ref{fig:mf_gain} shows the case in which the residual gains are correlated in both time and frequency. The mean SNR curve again remains close to the no-gain result and peaks near the expected bubble size. However, the realization-to-realization scatter is noticeably larger than in the time-only and frequency-only cases. This indicates that simultaneous correlations in time and frequency increase the variation in the recovered SNR, although the mean bubble scale remains close to the no-gain result.

Overall, the matched-filter technique remains robust for all three residual gain-correlation models considered here for the adopted residual gain errors of $\sigma_A=1\%$ and $\sigma_\phi=1^\circ$. We further performed tests with larger residual gain errors ($\sigma_A=10\%$ and $\sigma_\phi=5^\circ$), which show that the mean matched-filter response remains close to the no-gain result, although the uncertainty associated with different gain realizations increases and the recovered peak bubble size shows somewhat larger variations. This suggests that modest increases in the residual gain amplitude are unlikely to strongly affect matched-filter bubble detection, although we do not attempt to determine a precise tolerance threshold here. In contrast, as studied in the subsequent section, residual gain errors with the same parameters ($\sigma_A=1\%$ and $\sigma_\phi=1^\circ$) can severely hamper the detection of HI 21-cm images and the measurement of the power spectrum. 

The robustness of the matched-filter result is mainly due to the way the estimator combines the visibility data. The matched-filter estimator, as discussed in Section~\ref{sec:MF}, sums the visibility over all baselines and frequency channels. The bubble signal is well defined and is optimally combined over the entire observing session using a matched filter. Whereas the residual gain-error contribution does not exhibit the same coherent pattern across all baselines and frequencies. As a result, the residual gain-error contamination is averaged down in the summation. This reduces its impact on the recovered bubble scale. There is also additional averaging when $25$ independent $4$-h gain realizations are combined to obtain a total observation time of $100$-h.

The matched-filter results therefore indicate that the detection of an individual ionized bubble can remain robust for the residual gain errors considered here. 
The following section discusses the impact of residual gain errors on the reconstructed HI 21-cm images and the recovered power spectrum.   

\subsection{Impact of residual gain errors on HI 21-cm image}
\label{subsec:imaging_results}

Here, we first examine the effect of residual gain errors on HI 21-cm images. This provides a direct view of how calibration errors distort the 21-cm signal in the reconstructed images. We note that this effect has already been investigated in the context of 21-cm image analysis by \cite{Pal_2025}. They find that, in order to reliably recover the reionization history using images based on the Largest Cluster Statistics (LCS), residual gain calibration errors must be limited to a maximum of $\sim 0.02\%$.
However, the above study does not account for the possibility that residual gain errors may be correlated in time and frequency. In our study, we consider time-correlated, frequency-correlated, and joint time-frequency-correlated gain-error models to investigate their effects on the recovered 21-cm image and to compare these effects with those on the matched-filter detection of ionized regions.

Figure~\ref{fig:imaging_100h_gainonly} compares the original 21-cm brightness-temperature map with the corresponding reconstructed images obtained from the simulated SKA-Low observations in the absence of system noise. The top left panel shows the original simulated mean-subtracted 21-cm image at the central frequency channel. The top right panel shows the reconstructed image after foreground subtraction for the no residual gain case. The targeted ionized region is clearly seen at the centre of the field as a negative brightness-temperature feature. However, we find that pixels with very high \(\delta T_b\) values, exceeding \(50\) mK, are not fully recovered. This is because we consider only baselines shorter than \(2000\lambda\). The longer baselines are clipped. As a result, the reconstructed image is smoothed to some extent. This results in a loss of small-scale, high-\(\delta T_b\) features. 
The absence of the zero-spacing mode means that the reconstructed image traces the mean-subtracted differential brightness temperature ($\delta T_b - \bar{\delta T_b}$) rather than the absolute $21$-cm brightness temperature.

The bottom three panels of Fig. \ref{fig:imaging_100h_gainonly} show the reconstructed images in the presence of time-correlated (left), frequency-correlated (middle), and joint time-frequency-correlated (right) residual gain errors. In all three cases, the residual gain errors introduce strong noise-like structures across the reconstructed image. The targeted ionized region is completely buried in these noise-like structures and, therefore, no longer identifiable in the gain-corrupted images. In the absence of gain errors, the reconstructed \HI\ 21-cm brightness temperature spans approximately $-35$ to $58$~mK. In comparison, the reconstructed brightness temperature in the time-correlated gain-error case spans approximately $-890$ to $874$~mK, while the frequency-correlated case spans approximately $-329$ to $344$~mK. For the joint time-frequency-correlated model, the HI 21-cm brightness temperature in the reconstructed image ranges from approximately $-651$ to $573$~mK. We see that the frequency-correlated case introduces noise-like structures with the lowest amplitude.  

In this case, the residual gain errors for a particular antenna are correlated over $\sim \nu_{\rm corr}$ along the frequency axis. Therefore, artifacts produced due to this kind of residual gain errors are also correlated along frequency.  A portion of these smooth residuals is subsequently fitted out by the GPR foreground subtraction. In contrast, the time-correlated gain errors are uncorrelated across frequency channels and therefore introduce rapidly varying spectral structures. These structures are not captured by the smooth foreground model and consequently remain prominent after foreground subtraction. The joint time-frequency-correlated model contains correlations along both axes and produces an intermediate level of contamination between the time-only and frequency-only cases. The above results demonstrate that the level of residual gain errors must be several tens of times lower than that considered here to reliably detect ionized bubbles in \HI\ 21-cm images. However, the matched-filter technique should still be able to detect the bubbles with adequate significance in the presence of the level of residual gain contamination considered here.

Apart from the fiducial values of \(T_{\rm corr}=20\) s and \(\nu_{\rm corr}=1\) MHz, we also explore cases with different temporal and frequency correlation scales. Increasing \(T_{\rm corr}\) from $20$ s to $40$ s produces stronger noise-like structures in the reconstructed image. This makes the detection of the targeted ionized bubble more difficult. In contrast, increasing \(\nu_{\rm corr}\) from $1$ MHz to $5$ MHz results in a cleaner reconstructed image, as the smoother frequency-dependent residual gains are more efficiently removed by the GPR foreground subtraction. In contrast, we find that even in the presence of residual gain errors with higher temporal and frequency correlations, the matched-filter technique is still able to detect the targeted ionized bubble and reliably recover its size and the surrounding \HI\ neutral fraction. This further demonstrates the robustness of the matched-filter technique for detecting ionized bubbles even in the presence of relatively strong residual gain errors.

\subsection{Impact of residual gain errors on HI power spectrum}
\label{subsec:ps_results}
Here, we assess the impact of residual gain errors on the HI 21-cm power spectrum, which is one of the primary science goals of ongoing and upcoming EoR and Cosmic Dawn radio-interferometric experiments such as uGMRT, LOFAR, MWA, HERA, and SKA \citep{pal2021, Mertens_2020,Trott_2020,Abdurashidova_2022,Mellema_2013}.

Figure~\ref{fig:delta2_100h_gain} shows that residual gain errors alter the recovered power spectrum $\Delta^2(k)$ in a scale-dependent way. On large scales, $k\lesssim0.2$~Mpc$^{-1}$, the time-correlated gain model produces only a few per cent excess power relative to the no-gain reference. The frequency-correlated case shows a larger enhancement, typically at the level of several to a few tens of per cent. The joint time-frequency-correlated model produces the largest deviation on these scales, with the fractional excess generally reaching several tens of per cent and reaching about $2.7$ times the no-gain power near $k\simeq0.05$~Mpc$^{-1}$. Towards smaller spatial scales, the gain-induced contamination increases rapidly. Around $k\sim1$~Mpc$^{-1}$, the recovered power is approximately several hundred times larger than the no-gain reference for the time- and frequency-correlated models, while the joint time-frequency-correlated case reaches an enhancement of order $10^3$. Such biases can, in turn, affect the physical interpretation of the measured power spectrum.

\subsection{Bayesian parameter estimation}
\label{subsec:results_bayes}

We use a Bayesian framework to investigate how residual gain errors affect the recovery of the physical properties of the ionized bubble and the surrounding IGM. Following \cite{Ghara_2020}, we use the likelihood function developed for visibility-based matched-filter analyses.
The \HI\ 21-cm signal from the targeted bubble and the surrounding IGM is characterized by five parameters: the bubble radius ($R_f$), its angular coordinates relative to the antenna phase centre ($\theta_x$ and $\theta_y$), its position along the line of sight expressed as an offset from the central observing frequency ($\Delta\nu$), and the neutral hydrogen fraction of the surrounding IGM ($x_{\rm HI}$). We denote the parameter set as $\mu \equiv [R_f,\theta_x,\theta_y,\Delta\nu,x_{\rm HI}]$. For uniform priors, the logarithm of the likelihood is written as \citep{Ghara_2020},

\begin{equation}
     \begin{split}
    \log \Lambda(\mu) = \frac{1}{\sigma_{\rm rms}^2} \int d^2U \int d\nu \, \, \rho_B(\vec{U},\nu)
    \quad \left[ 2V(\vec{U},\nu) S_f^*(\vec{U},\nu;\mu)-\left| S_f(\vec{U},\nu;\mu) \right|^2 \right],
     \end{split} 
    \label{eq:log_likelihood}
\end{equation}
where $\rho_B(\vec{U},\nu)$ is the normalized baseline density satisfying $\int d^{2}U\int d\nu\,\rho_B(\vec{U},\nu)=1$, $\sigma_{\rm rms}$ denotes the system-noise rms, $V(\vec{U},\nu)$ represents the foreground-subtracted visibility data, and $S_f(\vec{U},\nu;\mu)$ is the matched-filter template corresponding to the parameter set $\mu$. In the gain-corrupted cases, the residual gain errors are included in the simulated visibility data, while the matched-filter template itself does not explicitly model the residual gain fluctuations. The posterior distributions, therefore, provide a direct measure of how residual calibration errors affect the recovered bubble and IGM parameters.

We explore the parameter space using the \emph{emcee} \citep{Foreman_Mackey_2013} Markov Chain Monte Carlo (MCMC) sampler with $50$ walkers. We perform the analysis for three cases: (i) no residual gain errors and no added system noise, (ii) joint time- and frequency-correlated residual gain errors without system noise, and (iii) the same gain-error model with system noise corresponding to $100$~h of SKA -Low \AAstar observation. 
The uniform priors adopted in the MCMC analysis are listed in Table~\ref{tab:priors}.

\begin{table}
\centering
\caption{Uniform prior distributions of the five parameters used in the MCMC analysis.}
\begin{tabular}{lcc}
\hline
Parameter & Range \\
\hline
Bubble radius ($R$) & $10 < R < 50$ Mpc \\
Position ($\theta_x$) & $-90 < \theta_x < 90$ arcmin \\
Position ($\theta_y$) & $-90 < \theta_y < 90$ arcmin \\
Frequency shift ($\Delta\nu$) & $-8 < \Delta\nu < 8$ MHz \\
Neutral fraction ($x_{\mathrm{HI}}$) & $0 < x_{\mathrm{HI}} \le 1$ \\
\hline
\end{tabular}
\label{tab:priors}
\end{table}

\begin{figure}
    \centering
    \includegraphics[width=0.8\linewidth]{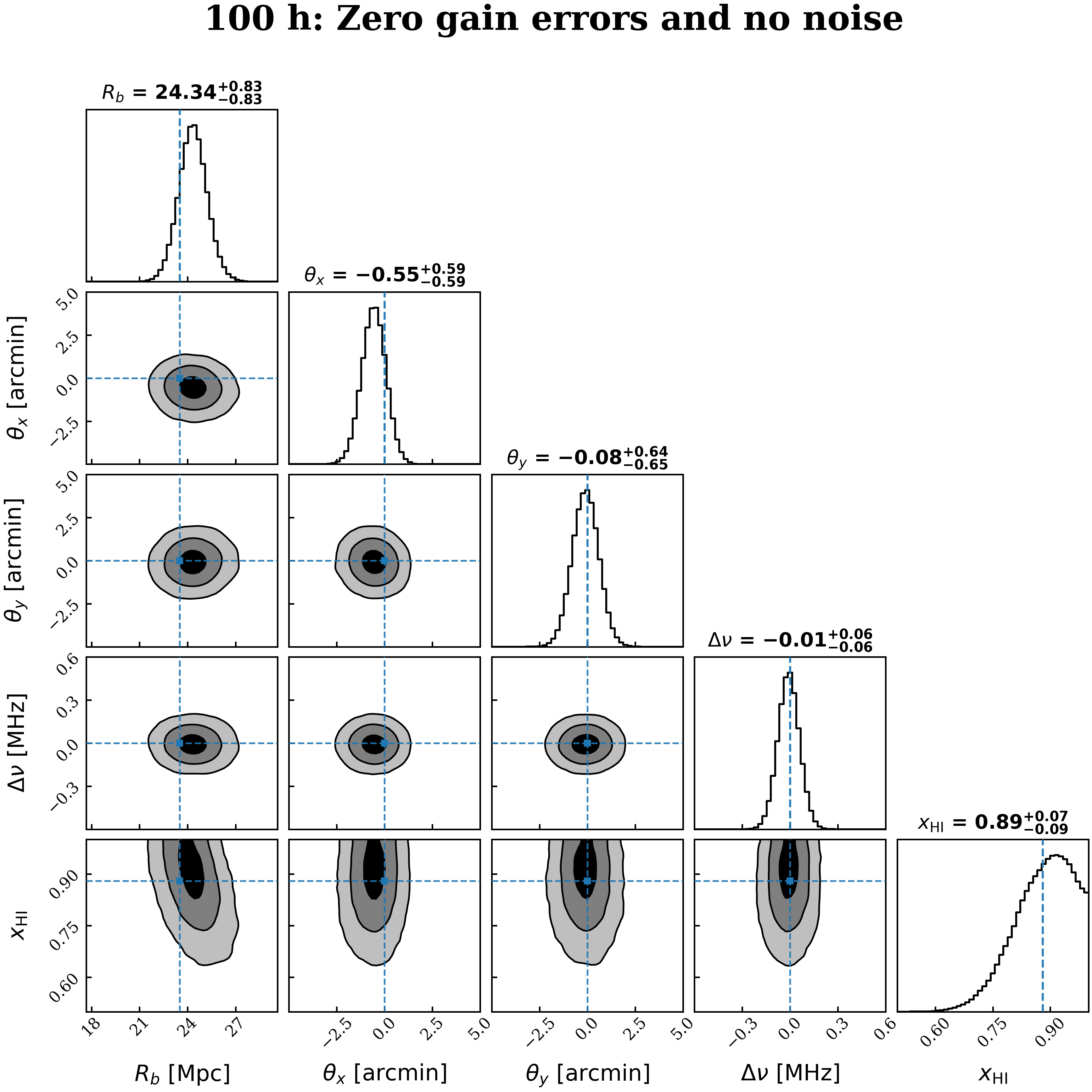}
    \caption{Posterior probability distributions of the ionized-bubble and surrounding-IGM parameters obtained from Bayesian analysis of the foreground-subtracted visibilities for $100$~h of SKA-Low \AAstar\ observations at $z=7.1$, in the absence of both residual gain errors and system noise. The diagonal panels show the one-dimensional marginalized distributions, while the contours represent the $1\sigma$, $2\sigma$, and $3\sigma$ credible regions. The blue dashed lines indicate the original parameter values.}
    \label{fig:mcmc_nogain}
\end{figure}

\begin{figure}
    \centering
    \includegraphics[width=0.8\linewidth]{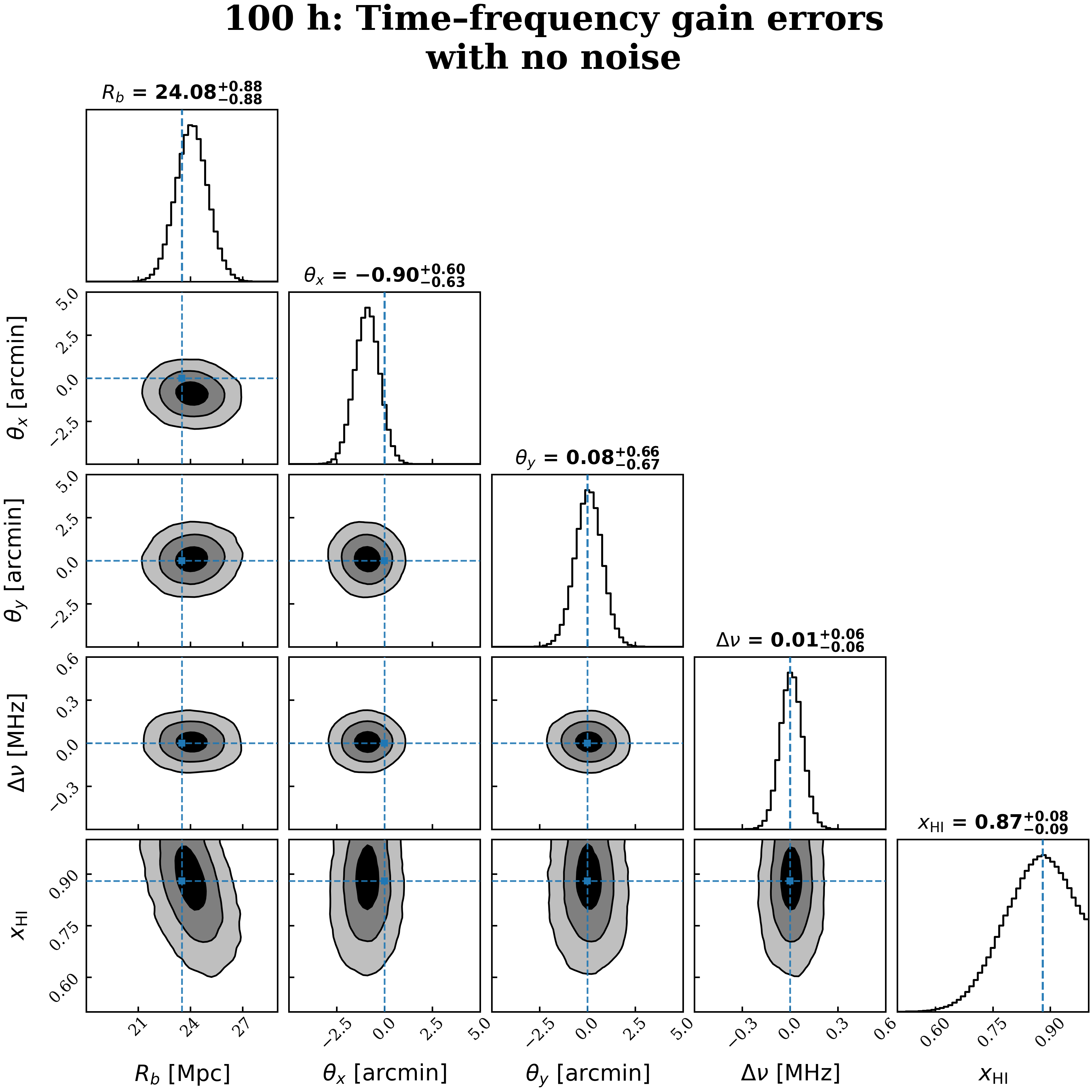}
    \caption{Posterior probability distributions of the ionized-bubble and surrounding-IGM parameters for $100$~h of SKA-Low \AAstar\ observations at $z=7.1$, in the presence of joint time- and frequency-correlated residual antenna gain errors but without system noise. The residual gain fluctuations have $\sigma_A=1\%$ and $\sigma_\phi=1^\circ$, with correlation scales $T_{\rm corr}=20$~s and $\nu_{\rm corr}=1$~MHz. The
    diagonal panels show the one-dimensional marginalized distributions, while the contours represent the $1\sigma$, $2\sigma$, and $3\sigma$ credible regions. The blue dashed lines indicate the original parameter values.}
    \label{fig:mcmc_joint_nonoise}
\end{figure}

\begin{figure}
    \centering
    \includegraphics[width=0.8\linewidth]{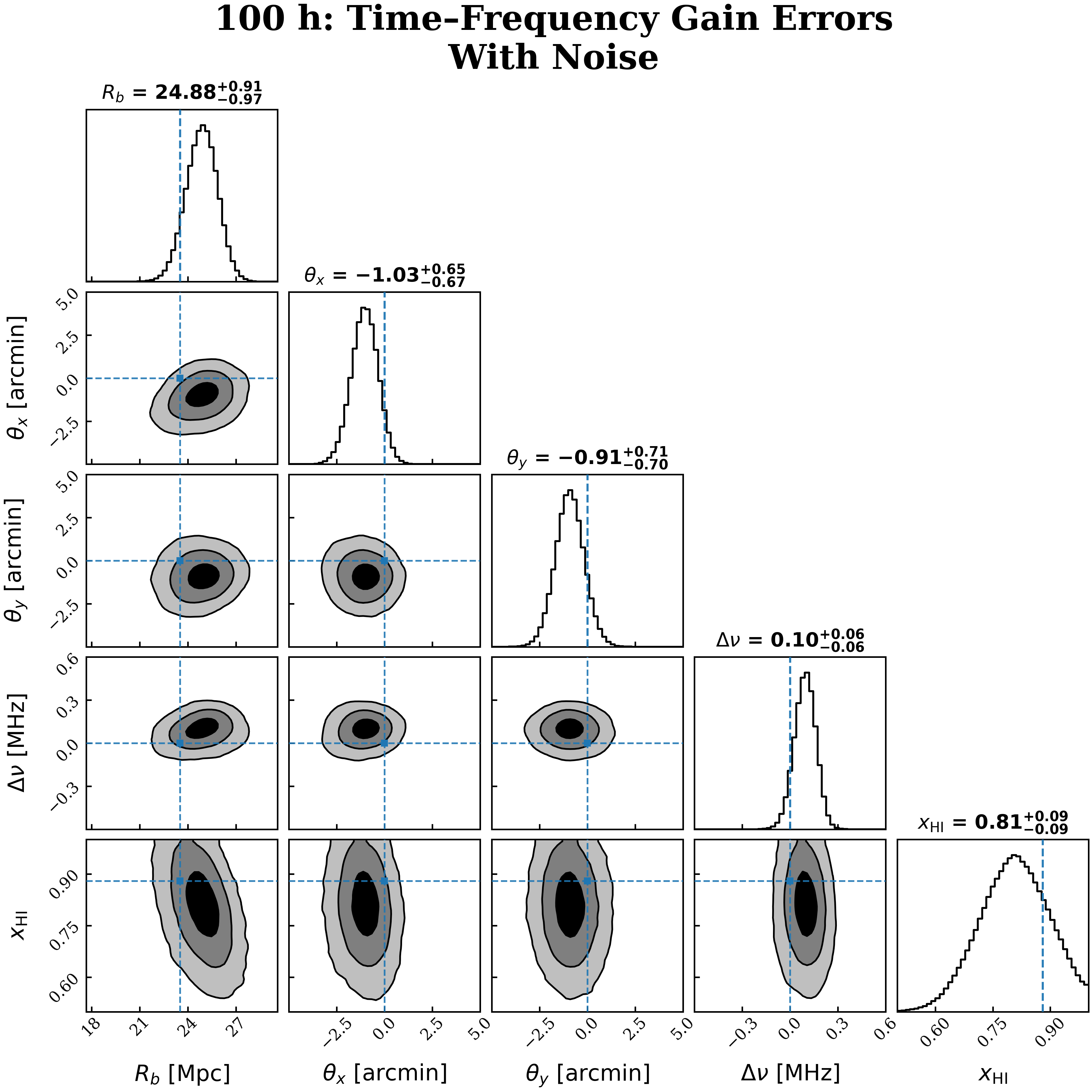}
    \caption{Posterior probability distributions of the ionized-bubble and surrounding-IGM parameters for $100$~h of SKA-Low \AAstar\ observations at $z=7.1$, in the presence of joint time- and frequency-correlated residual antenna gain errors and system noise corresponding to the $100$~h observation. The gain model is the same as in Figure~\ref{fig:mcmc_joint_nonoise}, with $\sigma_A=1\%$, $\sigma_\phi=1^\circ$, $T_{\rm corr}=20$~s, and $\nu_{\rm corr}=1$~MHz. The diagonal panels show the one-dimensional marginalized distributions, while the contours represent the $1\sigma$, $2\sigma$, and $3\sigma$ credible regions. The blue dashed lines indicate the original parameter values.}
    \label{fig:mcmc_joint_noise}
\end{figure}

Figures~\ref{fig:mcmc_nogain}--\ref{fig:mcmc_joint_noise} show the marginalized posterior distributions for the three cases, while the corresponding parameter constraints are summarized in Table~\ref{tab:mcmc_results}. We first consider the reference case without residual gain errors and without system noise. As shown in Figure~\ref{fig:mcmc_nogain}, all five parameters are well constrained. The recovered bubble radius is $R_{\rm b}=24.34^{+0.83}_{-0.83}$~Mpc, which is close to the original value of $23.5$~Mpc. The recovered angular positions are $\theta_x=-0.55^{+0.59}_{-0.59}$~arcmin and $\theta_y=-0.08^{+0.64}_{-0.65}$~arcmin, while the frequency offset is $\Delta\nu=-0.01^{+0.06}_{-0.06}$~MHz. The neutral hydrogen fraction is recovered as $x_{\rm HI}=0.89^{+0.07}_{-0.09}$, consistent with the original value $x_{\rm HI}=0.88$. Thus, in the absence of instrumental gain errors and system noise, the Bayesian matched-filter analysis recovers the original bubble and IGM parameters reasonably well.

\begin{table}[tbp]
\centering
\caption{Original and recovered parameters from the Bayesian MCMC analysis.
The joint gain model has $\sigma_A=1\%$, $\sigma_\phi=1^\circ$,
$T_{\rm corr}=20$~s, and $\nu_{\rm corr}=1$~MHz.}
\label{tab:mcmc_results}
\begingroup
\setlength{\tabcolsep}{4pt}
\renewcommand{\arraystretch}{1.35}
\begin{tabular}{@{}lcccc@{}}
\hline
Parameter & Original & \makecell{No gain errors\\No noise}
& \makecell{Joint gain errors\\No noise}
& \makecell{Joint gain errors\\With noise} \\
\hline
$R_{\rm b}$ [Mpc] & 23.5 & $24.34^{+0.83}_{-0.83}$ & $24.08^{+0.88}_{-0.88}$ & $24.88^{+0.91}_{-0.97}$ \\
$\theta_x$ [arcmin] & 0 & $-0.55^{+0.59}_{-0.59}$ & $-0.90^{+0.60}_{-0.63}$ & $-1.03^{+0.65}_{-0.67}$ \\
$\theta_y$ [arcmin] & 0 & $-0.08^{+0.64}_{-0.65}$ & $0.08^{+0.66}_{-0.67}$ & $-0.91^{+0.71}_{-0.70}$ \\
$\Delta\nu$ [MHz] & 0 & $-0.01^{+0.06}_{-0.06}$ & $0.01^{+0.06}_{-0.06}$ & $0.10^{+0.06}_{-0.06}$ \\
$x_{\rm HI}$ & 0.88 & $0.89^{+0.07}_{-0.09}$ & $0.87^{+0.08}_{-0.09}$ & $0.81^{+0.09}_{-0.09}$ \\
\hline
\end{tabular}
\endgroup
\end{table}

We next consider the effect of the joint time- and frequency-correlated residual gain errors, without system noise. The corresponding posteriors are shown in Figure~\ref{fig:mcmc_joint_nonoise}. The recovered parameters remain very close to those obtained for the no-gain reference. The bubble radius is recovered as $R_{\rm b}=24.08^{+0.88}_{-0.88}$~Mpc and the neutral fraction as $x_{\rm HI}=0.87^{+0.08}_{-0.09}$. The recovered angular position is $\theta_x=-0.90^{+0.60}_{-0.63}$~arcmin and $\theta_y=0.08^{+0.66}_{-0.67}$~arcmin, while the frequency offset is $\Delta\nu=0.01^{+0.06}_{-0.06}$~MHz. 
This comparison shows that the joint time- and frequency-correlated residual gains considered here introduce only a small additional change in the inferred parameters when system noise is absent. 

Finally, we consider the same joint time--frequency-correlated residual-gain model together with system noise corresponding to the $100$~h observation. The posterior distributions are shown in Figure~\ref{fig:mcmc_joint_noise}. The recovered bubble radius is $R_{\rm b}=24.88^{+0.91}_{-0.97}$~Mpc. The angular positions are recovered as $\theta_x=-1.03^{+0.65}_{-0.67}$~arcmin and $\theta_y=-0.91^{+0.71}_{-0.70}$~arcmin, while the frequency offset is $\Delta\nu=0.10^{+0.06}_{-0.06}$~MHz. The neutral hydrogen fraction is recovered as $x_{\rm HI}=0.81^{+0.09}_{-0.09}$. Compared with the gain-only case, the inclusion of system noise produces somewhat larger shifts in the recovered parameter values. It also produces a modest broadening of the uncertainty in posterior distributions for the recovered bubble parameters. 

Overall, the comparison between the three cases shows that the residual gain errors alone have only a modest effect on the Bayesian parameter recovery for the adopted gain model. The inclusion of system noise produces a larger change in some of the inferred parameters, particularly the neutral hydrogen fraction, although the posterior distributions remain well localized. These results indicate that the visibility-based matched-filter framework can be used not only to detect the ionized bubble but also to constrain its physical properties in the presence of residual calibration errors and observational system noise.

\section{Summary \& Discussion}
\label{sec:summary}

Detecting individual ionized regions around luminous sources during the epoch of reionization using the redshifted \HI\  21-cm signal provides a direct probe of the epoch. This approach also complements other statistical probes, such as measurements of the \HI\ 21-cm power spectrum and bispectrum. 
In this work, we have investigated how residual antenna gain errors affect matched-filter detection and recovery of ionized bubble and IGM parameters. We then compare these results with the detection of ionized regions using \HI\ 21-cm images and the recovered power spectrum. We have considered residual gain fluctuations that are correlated in time and frequency.

We have generated mock SKA-Low \AAstar\ observations of an ionized bubble at $z=7.1$, embedded in an IGM with $x_{\rm HI}=0.88$. The simulated visibilities contain the \HI\ 21-cm signal, astrophysical foregrounds, residual antenna gain errors, and system noise. We consider a total observing time of $100$~h, accumulated over $25$ independent $4$-h observing sessions. For the fiducial residual gain-error model, the rms amplitude and phase fluctuations are $\sigma_A=1\%$ and $\sigma_\phi=1^\circ$, respectively. Further, we consider three gain-correlation models: a time-correlated model with $T_{\rm corr}=20$~s, a frequency-correlated model with $\nu_{\rm corr}=1$~MHz, and a joint time--frequency-correlated model with $T_{\rm corr}=20$~s and $\nu_{\rm corr}=1$~MHz. 
We have also explored other values of the time and frequency correlation parameters. The foreground contribution, consisting of Galactic diffuse synchrotron radiation and extragalactic radio point sources, is removed using Gaussian Process Regression. Subsequently, the residual visibilities are analysed using the visibility-based matched-filter estimator. The residual visibilities are also used to construct images and estimate the power spectrum. These are then compared with the results obtained using the matched-filter technique.  

Our main result is that the matched-filter detection of the ionized bubble remains comparatively robust in the presence of the residual gain errors considered here. In the time-correlated,
frequency-correlated, and joint time-frequency-correlated cases, the matched-filter SNR continues to peak close to the actual bubble scale. Different gain realizations introduce some variation in the measured SNR, with the smallest variation seen for the time-correlated case and the largest uncertainty associated with the joint time-frequency-correlated model, but the recovered bubble scale remains relatively stable. The dominant effect of the residual gain errors on the matched-filter measurement is an increase in the uncertainty of the detection significance rather than a significant systematic shift in the recovered bubble size.

This robustness can be understood from the way the matched-filter estimator combines the visibility data. The estimator sums the visibilities over a large number of baselines and frequency channels. The 21-cm \HI\ signal associated with an ionized bubble embedded in an otherwise neutral hydrogen medium exhibits a regular, well-defined pattern. When multiplied by an appropriately matched filter, this signal adds coherently in the summation. In contrast, the contribution from residual gain errors does not maintain coherence across all baselines and frequency channels and is therefore partially averaged down in the summation. Further averaging occurs when the $25$ independent observing sessions are combined to form the full $100$~h observation. These effects reduce the impact of the residual gain errors on the recovered bubble scale.

In contrast, the residual gain errors have a much stronger effect on the reconstructed \HI\ 21-cm images and the recovered power spectrum. In the reconstructed images, the gain errors introduce strong noise-like structure. The targeted ionized bubble is completely buried within it and is not visually identifiable. The contamination is strongest for the time-correlated model, intermediate for the joint time-frequency-correlated model, and weakest for the frequency-correlated case. We find that the frequency-correlated residual gain errors are partially removed during the GPR foreground subtraction as they vary more smoothly with frequency. As a result, the frequency-correlated case produces less contamination in the reconstructed image than the time-correlated and joint time-frequency-correlated cases.
The corresponding power spectra also show additional scale-dependent power relative to the no-gain reference. The effect is relatively small on large spatial scales, but increases rapidly towards smaller scales. Around $k\sim1$~Mpc$^{-1}$, the recovered power is several hundred times larger than the no-gain result for the time- and frequency-correlated models, while the joint time-frequency-correlated case shows the largest contamination, with an enhancement of order $10^3$. 

We have further carried out a Bayesian parameter-estimation analysis using an MCMC framework in the presence of residual gain errors. The posterior distributions show that the bubble radius, its position, and the neutral hydrogen fraction of the surrounding IGM can still be constrained in the presence of both time- and frequency-correlated residual gain errors. The residual gains modify the uncertainty and, to some extent, the locations of the recovered posterior distributions, but useful and reliable constraints on the physical properties of the bubble and its surrounding IGM are still obtained for the $100$~h observations considered here.   

Taken together, our results demonstrate that individual ionized bubbles can remain detectable with high significance using visibility-based matched filtering for the time-correlated, frequency-correlated, and joint time-frequency-correlated residual gain error models considered here. At the same time, image reconstruction and power-spectrum measurements remain more susceptible to residual calibration structure. The matched-filter approach therefore provides a robust, promising, and complementary route for both detecting individual ionized regions and constraining their physical properties with future SKA-Low observations.

However, there are several limitations to the present analysis. We model the ionized region using a spherical matched-filter template, whereas realistic ionized bubbles may have more complex and asymmetric morphologies. A mismatch between the actual bubble shape and the assumed template could reduce the matched-filter SNR and affect the recovered bubble parameters. We also do not explicitly investigate the impact of frequency-dependent baseline migration across the observing band. Such migration can introduce additional chromatic structure in the gridded visibilities and may influence both foreground subtraction and the subsequent signal recovery. In addition, the residual gains considered here are direction-independent, and direction-dependent calibration effects are beyond the scope of the present analysis. Finally, our $100$~h observation assumes independent residual gain realizations between the $25$ individual $4$-h observing sessions. Residual calibration errors that remain correlated between different observing sessions may average down less efficiently and could therefore have a larger impact than considered here. Investigating these effects with more realistic bubble morphologies and calibration models will be important in future work.

\section*{Acknowledgements}
AM  acknowledges financial support from Council of Scientific and Industrial Research (CSIR) via  CSIR-SRF fellowships under grant no. 09/0096(13611)/2022-EMR-I. AM, KKD, SSA acknowledge financial support from ANRF
(Govt. of India) under the ARF program (File Number: ANRF/ARG/2025/004594/PS). IN acknowledges financial support from the DST-WISE Fellowship (DST/WISE-PhD/PM/2023/104), Govt. of India. SS acknowledges financial support by the Junior Research Fellowship of the University Grants Commission, Govt. of India under Ref. No. 231620034527. The computation work is performed using the facility procured through the financial support of DST-FIST program, Govt. of India provided to the Department of Physics, Jadavpur University vide sanction no. SR/FST/PS-1/2022/219(C).



\bibliographystyle{JHEP}
\bibliography{biblio_new}

\end{document}